\pdfoutput=1

\documentclass[11pt,twoside,a4paper,cmspaper,final,collab]{cms-tdr}
\def\svnVersion{465175P}\def\cmsCernNoTag{CERN-EP-2017-322}\def\cmsCernDate{\today}\def\cmsMessage{Published in Physical Review Letters as \href{http://dx.doi.org/10.1103/PhysRevLett.120.241801}{\doi{10.1103/PhysRevLett.120.241801.}}}

\begin{document}\cmsNoteHeader{SUS-17-006}

\hyphenation{had-ron-i-za-tion}
\hyphenation{cal-or-i-me-ter}
\hyphenation{de-vices}
\RCS$HeadURL: svn+ssh://svn.cern.ch/reps/tdr2/papers/SUS-17-006/trunk/SUS-17-006.tex $
\RCS$Id: SUS-17-006.tex 453819 2018-04-02 22:46:56Z wtford $

\newlength\cmsFigWidth
\ifthenelse{\boolean{cms@external}}{\setlength\cmsFigWidth{0.80\columnwidth}}{\setlength\cmsFigWidth{0.33\textwidth}}
\ifthenelse{\boolean{cms@external}}{\providecommand{\cmsLeft}{top\xspace}}{\providecommand{\cmsLeft}{left\xspace}}
\ifthenelse{\boolean{cms@external}}{\providecommand{\cmsRight}{bottom\xspace}}{\providecommand{\cmsRight}{right\xspace}}

\providecommand{\CLs}{\ensuremath{\text{CL}_{\text{S}}}\xspace}
\ifthenelse{\boolean{cms@external}}{\providecommand{\CL}{C.L.\xspace}}{\providecommand{\CL}{CL\xspace}}
\newcommand{\MHT}{\ensuremath{H_\mathrm{T}^{\text{miss}}}\xspace}
\newcommand\imini{\ensuremath{I}\xspace}
\providecommand{\mt}{\ensuremath{m_{\text{T}}}\xspace}
\newcommand{\mj}{\ensuremath{m_{\text{J}}}\xspace}

\cmsNoteHeader{SUS-17-006}
\title{Search for physics beyond the standard model in events with high-momentum Higgs bosons and missing transverse momentum in proton-proton collisions at \texorpdfstring{13\TeV}{13 TeV}}

\date{\today}

\abstract{
A search for physics beyond the standard model in events with one or more high-momentum Higgs bosons, \PH, decaying to pairs of \PQb quarks in association with missing transverse momentum is presented.
The data, corresponding to an integrated luminosity of 35.9\fbinv,
were collected with the CMS detector at the LHC in proton-proton
collisions at the center-of-mass energy $\sqrt{s}=13\TeV$. The analysis
utilizes a new \PQb quark tagging technique based on jet substructure to
identify jets from $\PH\to\PQb\PAQb$.
Events are categorized by the multiplicity of \PH-tagged jets,
jet mass, and the missing transverse momentum.
No significant deviation from standard model expectations is observed.
In the context of supersymmetry (SUSY), limits on the cross
sections of pair-produced gluinos are set, assuming
that gluinos decay to quark pairs, \PH (or \PZ), and the lightest SUSY particle, LSP, through an
intermediate next-to-lightest SUSY particle, NLSP.  With large mass
splitting between the NLSP and LSP, and 100\% NLSP branching fraction to \PH,
the lower limit on the gluino mass is found to be 2010\GeV.
}

\hypersetup{%
pdfauthor={CMS Collaboration},%
pdftitle={Search for physics beyond the standard model in events with high-momentum Higgs bosons and missing transverse momentum in proton-proton collisions at 13 TeV},%
pdfsubject={CMS},%
pdfkeywords={CMS, physics, supersymmetry, higgs, boosted}}

\maketitle

Primary motivations for building the CERN LHC~\cite{1748-0221-3-08-S08001} were
to determine the source of electroweak symmetry breaking and
to search for physics beyond the standard model (SM). In 2012, the first goal was
achieved with the discovery of the Higgs boson \PH by the ATLAS and CMS Collaborations~\cite{Aad:2012tfa,Chatrchyan:2012ufa,Aad:2015zhl}.
In this Letter, we exploit that discovery in a search
for events containing high-momentum Higgs bosons in conjunction with hadronic jets and
missing momentum transverse to the beam, \ptvecmiss.
Large $\ptmiss \equiv \abs{\ptvecmiss}$ can arise from the production of energetic weakly
interacting particles that escape detection.
A new particle of this type would be a candidate for weakly
interacting massive particle (WIMP) dark matter~\cite{Drees:2016pdg,Zwicky:1933gu,Rubin:1970zza}.
High-momentum Higgs bosons appear rarely
in SM processes, and would provide a unique signature of new physics.
Such a signature can arise in
a variety of models for physics beyond the SM, including extended electroweak
sectors~\cite{Dobrescu:2014fca,Dobrescu:2015asa}, extended Higgs sectors~\cite{Barman:2016kgt}, and
supersymmetry (SUSY)~\cite{Gori:2011hj,Kang:2015nga}.

The search presented here is based on 35.9\fbinv of proton-proton
(\Pp\Pp) collision data at $\sqrt{s}=13\TeV$ collected in 2016 by the CMS experiment.
High-momentum Higgs bosons are reconstructed in the leading $\PQb\PAQb$ decay channel in a regime in which the two jets
from the hadronization of the \PQb quarks overlap with each other.  They are identified with a recently
developed algorithm~\cite{BTV-16-002} that employs substructure techniques to large-radius jets.
In previous studies CMS~\cite{Chatrchyan:2013mya,Khachatryan:2014mma} and ATLAS~\cite{Aad:2012jva} have searched for
signatures with Higgs bosons, jets, and \ptmiss.
This Letter presents the first search for pairs of Lorentz-boosted Higgs bosons produced in association with jets and \ptmiss.

\begin{figure}[tbh]
  \centering
  \includegraphics[width=\cmsFigWidth]{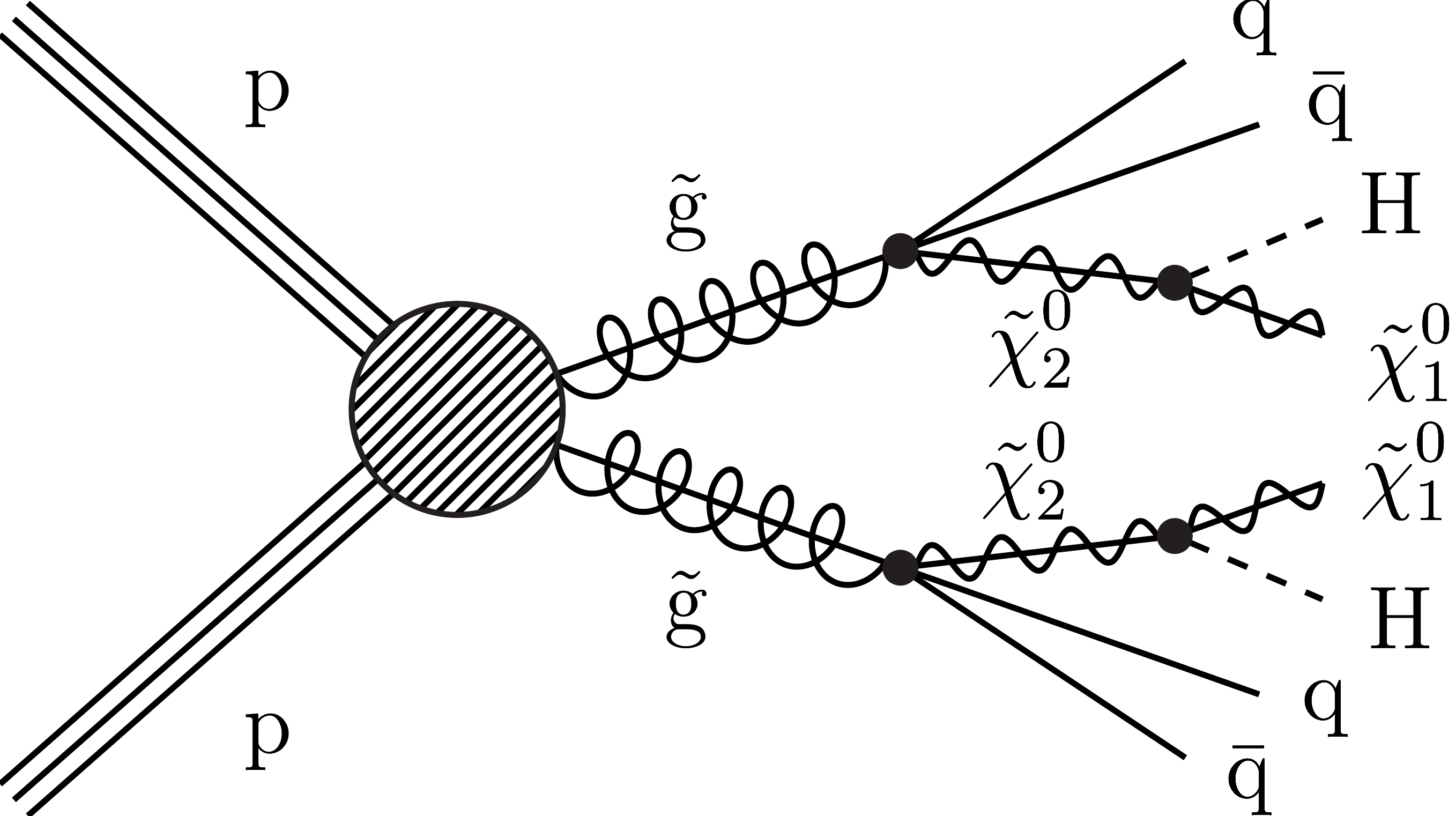}
  \caption{
Diagram for production of Higgs bosons via gluino pair production.  We also consider
channels in which a \PZ boson is substituted for \PH in one of the gluino decays.
    } \label{fig:results:SMSmodel}
\end{figure}

Supersymmetry~\cite{Ramond:1971gb,Golfand:1971iw,Neveu:1971rx,
Volkov:1972jx,Wess:1973kz,Wess:1974tw,Fayet:1974pd,Nilles:1983ge} is
a widely studied extension of the SM that posits for each SM particle a new particle,
called a superpartner,
with a spin that differs from that of its SM counterpart by a half unit.
Supersymmetry is attractive as a potential solution to the gauge hierarchy problem~\cite{Barbieri:1987fn}
that can help to explain the low mass of the Higgs boson without fine
tuning of the theory~\cite{Dimopoulos:1995mi,Barbieri:2009ev,Papucci:2011wy}.
The superpartners of quarks and gluons are squarks {\sQua}
and gluinos {\PSg}, respectively,
while neutralinos \PSGcz and charginos \PSGcpm
are mixtures of the superpartners of
the Higgs and electroweak gauge bosons.
In a process such as the simplified model
(SMS~\cite{bib-sms-2,bib-sms-3,bib-sms-4}) referred to as T5HH and
illustrated in Fig.~\ref{fig:results:SMSmodel}, gluinos are pair produced and decay into
a quark, antiquark, and \PSGczDt, where \PSGczDt is the second-lightest neutralino.  The \PSGczDt
decays into a Higgs boson and the lightest neutralino, \PSGczDo, which we take
to be the lightest SUSY particle (LSP) and represents the dark matter candidate. The results
of this search are interpreted in the context of this model and the alternate T5HZ,
in which the \PSGczDt branching fractions to $\PH\PSGczDo$ and $\PZ\PSGczDo$ are both 50\%,
with primary focus on the T5HH model.
We further assume a small \PSg{}--\PSGczDt mass splitting
and a light \PSGczDo, leading to events with energetic Higgs
bosons, large \ptmiss, and soft quark jets.

A detailed description of the CMS detector,
along with a definition of the coordinate system and
pertinent kinematical variables,
is given in Ref.~\cite{Chatrchyan:2008aa}.
Briefly,
a cylindrical superconducting solenoid with an inner diameter of 6\unit{m}
provides a 3.8\unit{T} axial magnetic field.
Within the solenoid volume
are a silicon pixel and strip tracker,
a lead tungstate crystal electromagnetic calorimeter (ECAL),
and a brass and scintillator hadron calorimeter (HCAL).
The tracking detectors cover the pseudorapidity range $\abs{\eta}<2.5$.
The ECAL and HCAL,
each composed of a barrel and two endcap sections,
cover $\abs{\eta}<3.0$.
Forward calorimeters extend the coverage to $\abs{\eta}<5.0$.
Muons are measured within $\abs{\eta}<2.4$ by gas-ionization detectors
embedded in the steel flux-return yoke outside the solenoid.
The detector is nearly hermetic,
permitting accurate measurement of~\ptmiss.

Individual particles are reconstructed
with the CMS particle-flow (PF)
algorithm~\cite{CMS-PRF-14-001},
which identifies them as photons,
charged hadrons, neutral hadrons, electrons, or muons.
Jets are defined by forming clusters of PF particles
using the anti-\kt jet algorithm~\cite{Cacciari:2008gp,Cacciari:2011ma}
with a distance parameter of~0.8 (AK8) and 0.4 (AK4). The jet energies are corrected for the nonlinear response of the
detector~\cite{Khachatryan:2016kdb} and to account for the expected contributions of neutral
particles from $\Pp\Pp$ interactions other than the one of interest (pileup)~\cite{Cacciari:2007fd}.
The quantity \ptvecmiss is reconstructed as the negative of the vector transverse momentum sum over all PF particles,
while \HT is the sum over AK4 jets of the magnitudes of their transverse momenta, \pt.
The jets for this summation are required to be within the tracker
volume and to have a minimum \pt of 30\GeV to suppress contributions from pileup.

The lepton content of events is used to characterize signal and
control samples.
We impose isolation requirements on electron and muon candidates
to suppress those arising from jets erroneously identified
as leptons, as well as genuine leptons from hadron decays.
The isolation criterion is based on the variable~$\imini$, defined as
the activity within a cone of radius $\sqrt{\smash[b]{(\Delta\phi)^2+(\Delta\eta)^2}}$
around the lepton direction divided by the lepton \pt.
Here activity is defined as the scalar \pt sum of charged hadron,
neutral hadron, and photon PF particles, corrected for the contributions
from pileup.
The radius of the cone is 0.2 for lepton $\pt<50\GeV$,
$10\GeV/\pt$ for $50\leq\pt\leq 200\GeV$,
and 0.05 for $\pt>200\GeV$.
The isolation requirement is $\imini<0.1$\,(0.2) for electrons (muons).

To recover electrons or muons that fail tight identification
requirements, and \PGt leptons via their one-prong hadronic decays, we
also make use of isolated charged tracks.
For these candidates we require that
the scalar \pt sum of all other charged-particle tracks
within a cone of radius 0.3 around the candidate track direction,
divided by the track~\pt,
be less than 0.2 if the track is identified
as a PF electron or muon
and less than 0.1 otherwise.
Isolated tracks are required to satisfy $\abs{\eta}<2.4$.

Candidates for $\PH\to\PQb\PAQb$ jets are identified with a heavy-flavor tagging algorithm designed to
identify a pair of \PQb quarks clustered into a single AK8 jet~\cite{BTV-16-002}.
The algorithm resolves the decay chains of the two \PQb hadrons and associates secondary vertices along the decay directions,
and then computes the likelihood that a jet contains two \PQb hadrons.
The jet pruning algorithm described in Ref.~\cite{JetSub:Prune} is used to
improve the jet mass resolution for $\PH\to\PQb\PAQb$ candidates.

The selection of events for analysis begins with the trigger described in Ref.~\cite{Khachatryan:2016bia}.
For this analysis,
signal event candidates were recorded
by requiring \ptmiss and the magnitude \MHT of the vector \pt sum of jets, both computed at the trigger level,
to exceed thresholds that varied between
100 and 120\GeV depending on the LHC instantaneous luminosity.
The efficiency of this trigger,
which exceeds 98\% for events satisfying the selection
criteria described below,
is measured in data and is taken into account in the analysis.
Additional triggers,
requiring the presence of charged leptons,
photons, or minimum values of \HT,
are used to select samples, described below, employed in the evaluation of backgrounds.

Candidates for signal events are characterized by jets of large angular radius containing
a pair of \PQb quarks from the decays of Lorentz-boosted Higgs bosons, accompanied by \ptmiss from escaping
LSPs. They are required to have no isolated leptons, but
we impose no requirements on the number of additional jets in the event.
The specific requirements that define the search sample are:
$\ptmiss>300\GeV$, $\HT>600\GeV$, and at least two AK8 jets with
$\pt>300\GeV$ and mass \mj between 50 and 250\GeV. We exclude events with either a muon or an
electron with $\pt>10\GeV$, or an isolated
track with $m_T<100\GeV$ and $\pt>10$\,(5)\GeV for hadronic (leptonic)
tracks.  Here \mt is the transverse mass~\cite{Arnison:1983rp}
evaluated from the \ptvecmiss and isolated-track \pt vectors.
The isolated track requirement serves to improve the efficiency for suppressing background from leptonic \PW\ decays.
To suppress events containing apparent \ptmiss caused by mismeasurement of the jet energies,
we further impose thresholds on the azimuthal angles between the \ptvecmiss
vector and those of the (up to) four leading-\pt AK4 jets, $\Delta\phi_{1,2,3,4} > 0.5,0.5,0.3,0.3$.
For enhanced sensitivity to diverse signal models, events are categorized into three ranges
of \ptmiss{}: 300--500, 500--700, and $>$700\GeV.

\begin{figure}[tbh]
  \centering
  \includegraphics[width=0.49\textwidth]{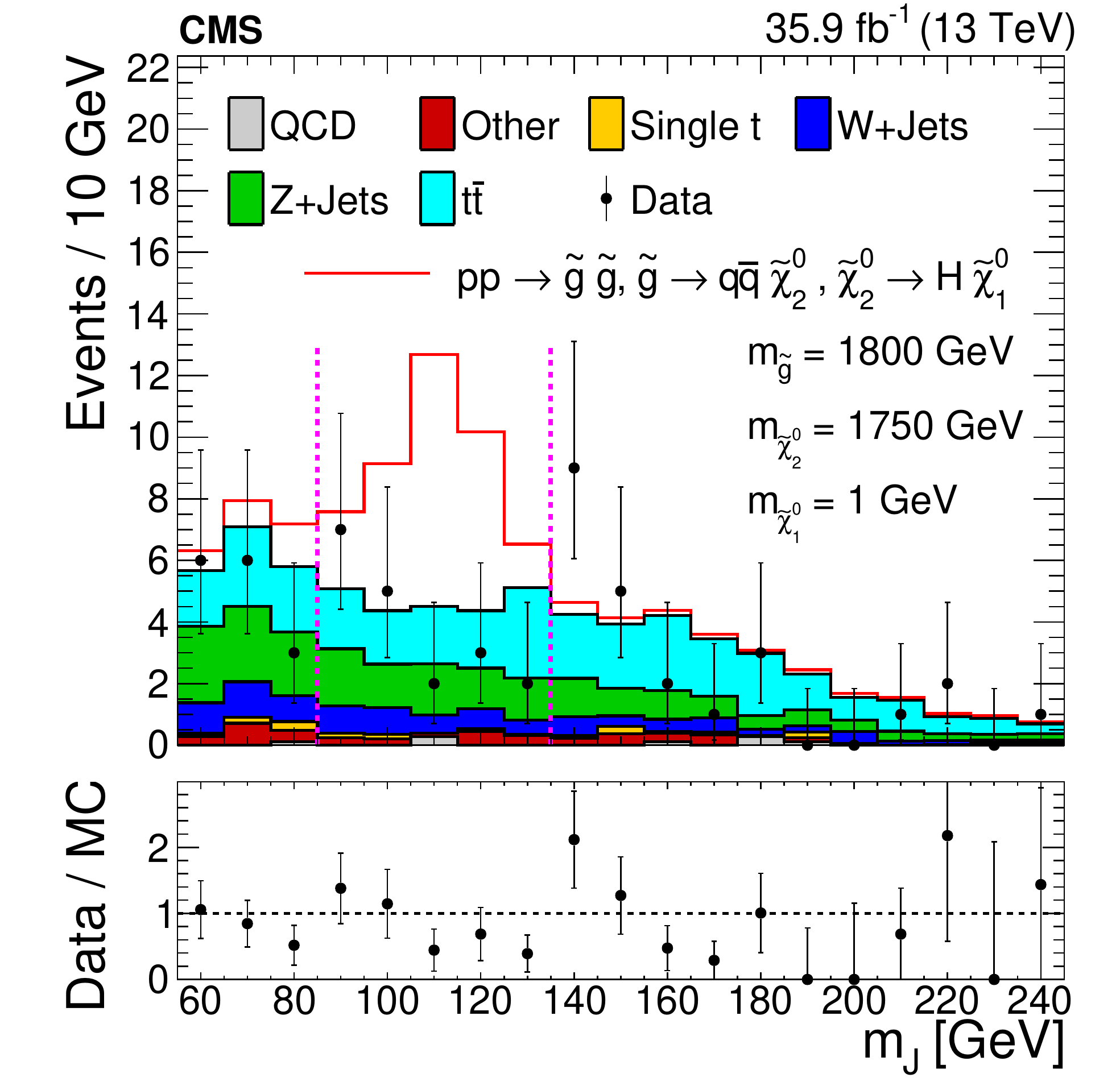}
  \caption{Observed and expected distributions of the leading-\pt jet mass for selected 1H and 2H
  	events with $\ptmiss > 300\GeV$.
        The subleading jet is required to have \mj within the signal window
denoted by vertical dashed magenta lines.
The yields from simulation are scaled to the
prediction based on control samples in data, in the signal mass window.
A representative signal is shown
stacked on top of the backgrounds.
	The bottom panel shows the ratio of the observed
  	to SM-expected yields.
    } \label{fig:event_reco:J1M}
\end{figure}

Considering the two leading-\pt AK8 jets, events are categorized as 0H, 1H, or 2H according to the number
of these jets that have a double-b discriminator value greater than 0.3 (\PH-tagged jets).
For true Higgs boson decays the efficiency of this requirement is 50--80\% per AK8 jet depending on the
jet \pt, with the maximum around $500\GeV$, dropping off to the lower value around 2\TeV.
Jets are further categorized by \mj, with
the Higgs signal region encompassing the range 85--135\GeV, for which the efficiency
per jet is $\sim$80\%. The remaining mass regions, 50--85
and 135--250\GeV, serve as sidebands.
The signal region $\text{A}_{1}$ ($\text{A}_{2}$) is defined as the class of
1H (2H) events in which both jets lie within
the signal mass window.  Distributions of \mj for the leading-\pt jet in 1H and 2H events are shown
in Fig.~\ref{fig:event_reco:J1M}, for the observed and simulated events in which the subleading
AK8 jet lies within the signal mass window.  Here the yields from simulation are scaled to the
prediction based on control samples in data, described below, in this mass window.
For the T5HH SUSY model the efficiency for selection of events in
the signal regions is 9--15\% for $m(\PSg)>1200\GeV$, increasing with $m(\PSg)$.

The \mj resolution does not permit clean separation of the \PH and \PZ boson peaks.  The chosen
signal window optimizes the selection of \PH bosons in the absence of \PZ.  As noted
previously we treat both processes as potential signal, and the likelihood fit described below
accounts for any signal population in the control regions.

Simulated event samples for SM background processes are used to determine correction factors, typically near
unity, that are used in conjunction with observed event yields in control regions to determine the SM
background contribution in the signal regions.  The production of \ttbar, \PW, \PZ, and quantum
chromodynamics (QCD) multijet events is simulated with the Monte Carlo (MC)
generator \MGvATNLO~2.2.2~\cite{Alwall:2014hca}, with parton distribution functions (PDFs) taken from NNPDF
3.0~\cite{Ball:2014uwa}.  A detailed description of the simulated SM background samples is given in
Ref.~\cite{Sirunyan:2017cwe}. The detector simulation is performed
with \GEANTfour~\cite{Agostinelli:2002hh}.  Simulated event samples for SUSY signal models, used to determine
the selection efficiency for signal events, are generated with \MGvATNLO with up to two additional
partons at leading order accuracy; they are normalized to cross sections computed to next-to-leading order
(NLO) plus next-to-leading logarithmic (NLL) accuracy, based on Ref.~\cite{Borschensky:2014cia}.

The signal efficiencies from simulation are corrected for the modeling of initial-state radiation as
measured in a data control sample~\cite{Sirunyan:2017cwe},
the double-\PQb tagging efficiency~\cite{BTV-16-002},
and the \mj resolution observed in data.  Systematic uncertainties associated with these
corrections are taken into account, as well as those arising from the determination of luminosity,
trigger efficiency, PDFs, jet energy scale and resolution, isolated track
veto efficiency, renormalization and factorization scales~\cite{Catani:2003zt,Cacciari:2003fi},
and predicted yields from simulation due to limited sample sizes. The largest
uncertainties are associated with the modeling
of the double-b tagging efficiency (6\%) and the mass resolution (1--15\%).

\label{sec:background}
Dominant SM backgrounds arise from events containing jets
misidentified as Higgs bosons in conjunction with \PW\ or \PZ\ bosons,
which may originate from top quark decays,
that decay to final states with neutrinos, yielding large \ptmiss.
Multijet events in which jets are undermeasured can also give large \ptmiss;
these backgrounds are highly suppressed by the Higgs boson identification
requirements. All backgrounds are estimated from control regions in the data.

The SM backgrounds are estimated by simultaneously extrapolating yields from
the 0H to the 1H and 2H \PH-tag multiplicity regions,
and from the \mj sideband to the signal window.
Events are assigned to the \mj sideband
if one or both of the leading-\pt jets lie outside
the signal window.
Altogether we define four control regions:  1H and 2H events in the \mj sidebands,
denoted $\text{B}_1$ and $\text{B}_2$, respectively; 0H events in the \mj signal window,
denoted C; and 0H events in the \mj sidebands, denoted D.
Each control region is split into three \ptmiss bins, corresponding to
those defined for the signal regions.  Based on the observed yields in these regions 
within the search sample,
the total background is estimated as\\
\begin{equation}
\mathcal{A}_{1,2} = N(\text{B}_{1,2}) \frac{N(\text{C})}{N(\text{D})} \kappa_{1,2},
\label{eq:abcdkappa}
\end{equation}
where the subscript indicates the number of double-b tagged jets,
$\mathcal{A}_{1,2}$ is the predicted yield in the $\text{A}_{1,2}$ signal region,
$N$ is the population of the indicated control region, and $\kappa_{1,2}$
is a correction factor used to account for any correlations between the \PH-tag and \mj variables.
While $\text{B}_{1,2}$, C, and D yields are taken directly
from data, $\kappa_{1,2}$ is computed from simulation, corrected for observed
discrepancies between data and simulation.

To obtain the corrections to $\kappa_{1,2}$ we compare data with simulation in
auxiliary samples,
defined to be orthogonal to the search sample, that are
enriched in the SM backgrounds expected in the signal 
region:  a single-lepton sample dominated by top quark and W boson production, a
sample of single-photon plus jets events serving as proxy for invisibly
decaying \PZ bosons \cite{Sirunyan:2017cwe}, and a sample selected by inverting the $\Delta\phi$
requirement that contains predominantly QCD multijet events.  
The auxiliary samples satisfy the same requirement in \HT and contain the same
control and signal regions, ($\text{B}_{1,2}$, C, D) and $\text{A}_{1,2}$, as the search
sample.  
Scale factors given by ratios of the yields in data divided by those from
simulation in these auxiliary samples, typically ranging in value
from 0.5 to 2.0, are then applied to the yields
of each of the simulated SM backgrounds, before they are combined to obtain the
total background yields in the signal and control regions of the search sample.
The yields from corrected simulation are found to be statistically
compatible with the data in the control regions.
From these corrected MC yields
we compute $\kappa_{1,2}$ via Eq.~\ref{eq:abcdkappa}, for each \ptmiss bin;
the values are given in Table~\ref{tab:DataPred} below.

Systematic uncertainties in the background prediction enter through
the factors $\kappa_{1,2}$.  These include contributions from the uncertainties
in the relative populations of the SM background processes, the
yield statistics and simulation self-consistency in the auxiliary samples,
the \ptmiss dependence of the scale factors where \ptmiss regions are
combined to reduce statistical uncertainties, and the self-consistency
of the method as applied to the simulated data.

The values of the $\kappa$ factors with their uncertainties for each
of the signal regions appear in Table~\ref{tab:DataPred}, along with
the final background yield predictions, and the yields observed in the
data.  The observations are statistically compatible with those
expected from the SM backgrounds, and thus we find no evidence for
processes outside the SM.

\begin{table}[htbp!]
\topcaption{Correction factors, predicted SM background yields, and observed yields,
for the signal regions $A_{N_\PH}$. The
uncertainties in the predictions include both statistical and systematic contributions.  }
\centering
\begin{scotch}{rcccr}
$N_\PH$ & \ptmiss (\GeVns{}) & $\kappa$ & Predicted & Observed\\
\hline
1 & $300-500$ & $0.98 \pm 0.11$ & $17.7 \pm 3.8$        & 15 \\
1 & $500-700$ & $0.86 \pm 0.16$ & $3.4  \pm 1.5$        & 2 \\
1 & $>$700    & $0.86 \pm 0.17$ & $0.61  \pm 0.45$        & 1 \\
2 & $300-500$ & $0.73 \pm 0.14$ & $1.52  \pm 0.57$        & 1 \\
2 & $500-700$ & $0.43 \pm 0.12$ & $0.09  \pm 0.08$      & 0 \\[0.3ex]
2 & $>$700    & $0.62 \pm 0.30$ & $0.09 ^{+0.11}_{-0.09}$ & 0\\
\end{scotch}
\label{tab:DataPred}
\end{table}

We compute upper limits on the gluino pair-production cross section using a maximum-likeli\-hood fit
in which the free parameters are the signal strength $\mu$, 
the Poisson means of the total expected yields from SM backgrounds in each of the $\text{B}_{1,2}$, C,
and D regions, and $\kappa_{1,2}$.
The $\kappa_{1,2}$ parameters are constrained with a Gaussian prior to the expected
values, with their statistical and systematic uncertainties.
The signal model in the fit accounts for the populations of control as well as signal regions.
Additional nuisance parameters
account for systematic uncertainties in the yields predicted by the signal model.

\begin{figure}[tbh]
  \centering
  \includegraphics[width=0.49\textwidth]{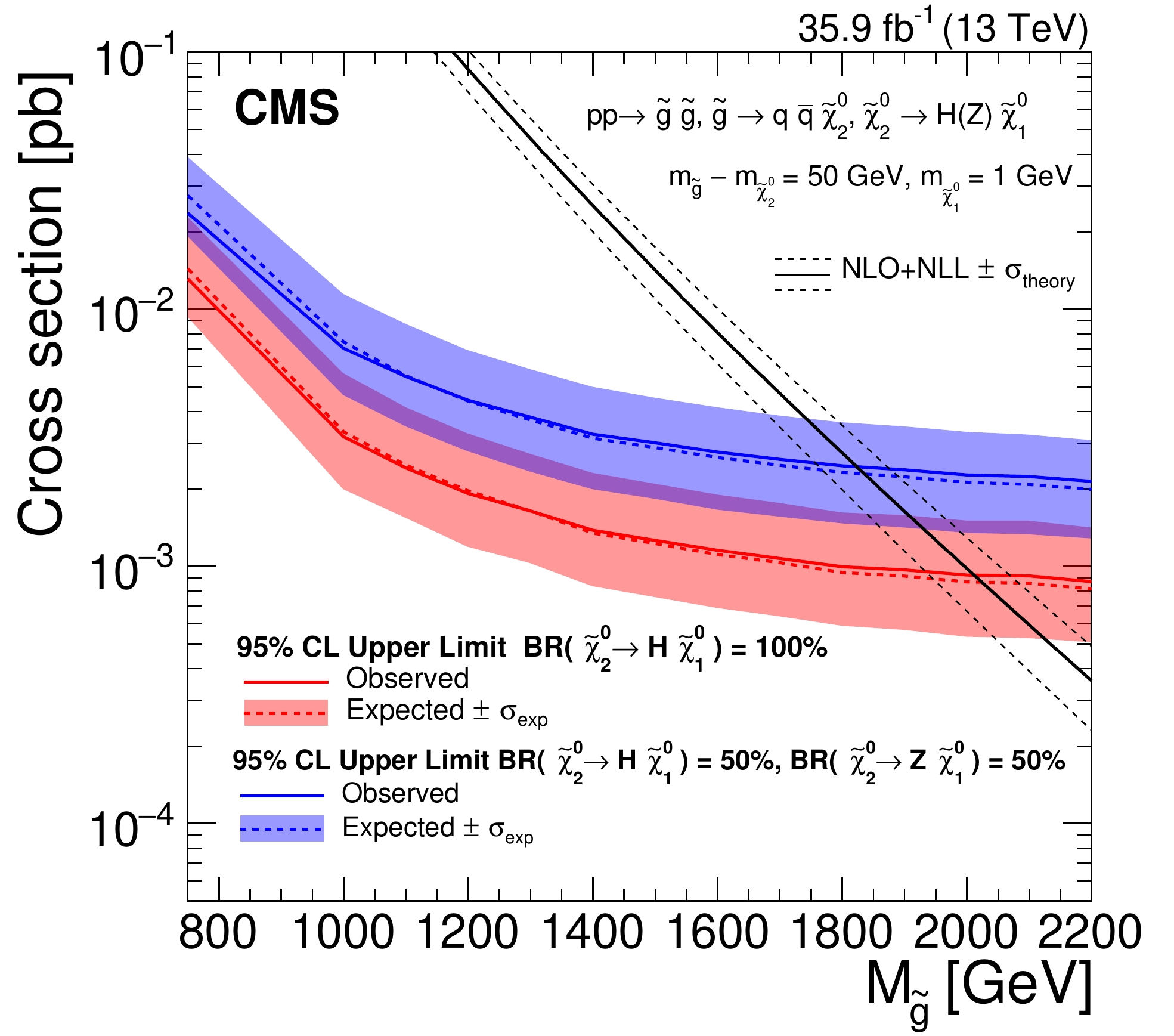}
  \caption{Observed and expected cross section upper bounds at 95\%~\CL for the T5HH and T5HZ
    models. The solid and dashed black lines show the SMS gluino-gluino production cross section with its uncertainty. The solid red (blue) line shows the observed limit for the T5HH (T5HZ) model; for each the like-colored dashed line and shaded band show the expected limit and the range associated with the experimental uncertainties.
    } \label{fig:results:interp}
\end{figure}

We evaluate 95\% confidence level (\CL{}) upper limits based on the asymptotic form of a likelihood ratio test
statistic~\cite{Cowan:2010js}, in
conjunction with the \CLs criterion described in Refs.~\cite{bib-cls,Junk1999,cms-note-2011-005}.  The test statistic is
$q(\mu) = -2 \ln(\mathcal{L}_{\mu}/\mathcal{L}_{\text{max}})$, where $\mathcal{L}_{\text{max}}$ is the maximum likelihood determined
by allowing all parameters, including
$\mu$, to vary, and $\mathcal{L}_{\mu}$ is the maximum likelihood for fixed $\mu$.
Expected and observed 95\% \CL upper limits, and the predicted gluino pair-production cross sections, are shown in
Fig.~\ref{fig:results:interp} for two choices of the \PSGczDt decay branching fractions,
taking $m(\PSGczDo)=1\GeV$ and $m(\PSg)-m(\PSGczDt)=50\GeV$.
That is, we choose a model with a light LSP and a compressed spectrum
for the heavy SUSY particles, thereby ensuring a Lorentz-boosted topology.

In summary, this Letter has presented a search for production of energetic
Higgs bosons in conjunction with large missing transverse momentum in
proton-proton collisions.  Higgs bosons with transverse momentum in the
range 300\GeV to about 2\TeV are reconstructed as wide-cone jets with
substructure indicative of the decay of the Higgs boson to a pair of b
quarks.  Background from standard model processes is estimated from
data control regions.  The observed event yields are found to be
statistically compatible with these backgrounds.

The results are broadly applicable to models leading to signatures with energetic Higgs
bosons and missing momentum.  Here they are interpreted in the context
of a simplified model of supersymmetry in which gluinos are
pair produced and subsequently decay into several quarks, a Higgs or \PZ
boson, and the lightest supersymmetric particle, a neutralino \PSGczDo.
Gluinos with masses below 2010\,(1825)\GeV are excluded under the assumption of
a large mass splitting between the next-to-lightest and lightest supersymmetric particle
and that the branching fraction of $\PSGczDt\to \PH\PSGczDo$ is 100\% (50\%).
These are the first limits for pair production of gluinos measured
in these decay channels.

\begin{acknowledgments}
We congratulate our colleagues in the CERN accelerator departments for the excellent performance of the LHC and thank the technical and administrative staffs at CERN and at other CMS institutes for their contributions to the success of the CMS effort. In addition, we gratefully acknowledge the computing centers and personnel of the Worldwide LHC Computing Grid for delivering so effectively the computing infrastructure essential to our analyses. Finally, we acknowledge the enduring support for the construction and operation of the LHC and the CMS detector provided by the following funding agencies: BMWFW and FWF (Austria); FNRS and FWO (Belgium); CNPq, CAPES, FAPERJ, and FAPESP (Brazil); MES (Bulgaria); CERN; CAS, MoST, and NSFC (China); COLCIENCIAS (Colombia); MSES and CSF (Croatia); RPF (Cyprus); SENESCYT (Ecuador); MoER, ERC IUT, and ERDF (Estonia); Academy of Finland, MEC, and HIP (Finland); CEA and CNRS/IN2P3 (France); BMBF, DFG, and HGF (Germany); GSRT (Greece); OTKA and NIH (Hungary); DAE and DST (India); IPM (Iran); SFI (Ireland); INFN (Italy); MSIP and NRF (Republic of Korea); LAS (Lithuania); MOE and UM (Malaysia); BUAP, CINVESTAV, CONACYT, LNS, SEP, and UASLP-FAI (Mexico); MBIE (New Zealand); PAEC (Pakistan); MSHE and NSC (Poland); FCT (Portugal); JINR (Dubna); MON, RosAtom, RAS, RFBR and RAEP (Russia); MESTD (Serbia); SEIDI, CPAN, PCTI and FEDER (Spain); Swiss Funding Agencies (Switzerland); MST (Taipei); ThEPCenter, IPST, STAR, and NSTDA (Thailand); TUBITAK and TAEK (Turkey); NASU and SFFR (Ukraine); STFC (United Kingdom); DOE and NSF (USA).
\end{acknowledgments}

\bibliography{auto_generated}

\providecommand{\href}[2]{#2}\begingroup\raggedright\begin{thebibliography}{10}%
\makeatletter
\providecommand{\hrefCMSnoop }[0]{\@secondoftwo}%
\makeatother
\providecommand{\doi}{\texttt{doi:}\begingroup \urlstyle{tt}\Url}

\bibitem{1748-0221-3-08-S08001}
\hrefCMSnoop {}{L.~Evans and P.~Bryant, ``{LHC} machine'',} \textit{ JINST}
  \textbf{ 3} (2008) S08001,
  \href{http://dx.doi.org/10.1088/1748-0221/3/08/S08001}{\doi{10.1088/1748-0221/3/08/S08001}}.

\bibitem{Aad:2012tfa}
\hrefCMSnoop {}{{ATLAS Collaboration}, ``{Observation of a new particle in the
  search for the Standard Model Higgs boson with the ATLAS detector at the
  LHC}'',} \textit{ Phys. Lett. B} \textbf{ 716} (2012) 1,
  \href{http://dx.doi.org/10.1016/j.physletb.2012.08.020}{\doi{10.1016/j.physletb.2012.08.020}},
\href{http://www.arXiv.org/abs/1207.7214}{\texttt{arXiv:1207.7214}}.

\bibitem{Chatrchyan:2012ufa}
\hrefCMSnoop {}{{CMS Collaboration}, ``{Observation of a new boson at a mass of
  125 GeV with the CMS experiment at the LHC}'',} \textit{ Phys. Lett. B}
  \textbf{ 716} (2012) 30,
  \href{http://dx.doi.org/10.1016/j.physletb.2012.08.021}{\doi{10.1016/j.physletb.2012.08.021}},
\href{http://www.arXiv.org/abs/1207.7235}{\texttt{arXiv:1207.7235}}.

\bibitem{Aad:2015zhl}
\hrefCMSnoop {}{{ATLAS and CMS Collaborations}, ``Combined measurement of the
  {H}iggs boson mass in $pp$ collisions at $\sqrt{s}=7$ and 8 {TeV} with the
  {ATLAS} and {CMS} experiments'',} \textit{ Phys. Rev. Lett.} \textbf{ 114}
  (2015) 191803,
  \href{http://dx.doi.org/10.1103/PhysRevLett.114.191803}{\doi{10.1103/PhysRevLett.114.191803}},
\href{http://www.arXiv.org/abs/1503.07589}{\texttt{arXiv:1503.07589}}.

\bibitem{Drees:2016pdg}
\hrefCMSnoop {}{{Particle Data Group}, C.~Patrignani {et~al.}, ``{Review of
  Particle Physics}'',} \textit{ Chin. Phys. C} \textbf{ 40} (2016) 100001,
  \href{http://dx.doi.org/10.1088/1674-1137/40/10/100001}{\doi{10.1088/1674-1137/40/10/100001}}.
See section 26, {Dark Matter}.

\bibitem{Zwicky:1933gu}
\hrefCMSnoop {}{F.~Zwicky, ``Die rotverschiebung von extragalaktischen
  nebeln'',} \textit{ Helv. Phys. Acta} \textbf{ 6} (1933) 110,
\href{http://dx.doi.org/10.5169/110267}{\doi{10.5169/110267}}.

\bibitem{Rubin:1970zza}
\hrefCMSnoop {}{V.~C. Rubin and W.~K. Ford~Jr, ``Rotation of the {A}ndromeda
  nebula from a spectroscopic survey of emission regions'',} \textit{
  Astrophys. J.} \textbf{ 159} (1970) 379,
\href{http://dx.doi.org/10.1086/150317}{\doi{10.1086/150317}}.

\bibitem{Dobrescu:2014fca}
\hrefCMSnoop {}{B.~A. Dobrescu and C.~Frugiuele, ``{Hidden GeV-scale
  interactions of quarks}'',} \textit{ Phys. Rev. Lett.} \textbf{ 113} (2014)
  061801,
  \href{http://dx.doi.org/10.1103/PhysRevLett.113.061801}{\doi{10.1103/PhysRevLett.113.061801}},
\href{http://www.arXiv.org/abs/1404.3947}{\texttt{arXiv:1404.3947}}.

\bibitem{Dobrescu:2015asa}
\hrefCMSnoop {}{B.~A. Dobrescu, ``Leptophobic boson signals with leptons, jets
  and missing energy'',} (2015).
\href{http://www.arXiv.org/abs/1506.04435}{\texttt{arXiv:1506.04435}}.

\bibitem{Barman:2016kgt}
\hrefCMSnoop {}{R.~K. Barman, B.~Bhattacherjee, A.~Chakraborty, and
  A.~Choudhury, ``{Study of MSSM heavy Higgs bosons decaying into charginos and
  neutralinos}'',} \textit{ Phys. Rev. D} \textbf{ 94} (2016) 075013,
  \href{http://dx.doi.org/10.1103/PhysRevD.94.075013}{\doi{10.1103/PhysRevD.94.075013}},
\href{http://www.arXiv.org/abs/1607.00676}{\texttt{arXiv:1607.00676}}.

\bibitem{Gori:2011hj}
\hrefCMSnoop {}{S.~Gori, P.~Schwaller, and C.~E.~M. Wagner, ``{Search for Higgs
  bosons in supersymmetric cascade decays and neutralino dark matter}'',}
  \textit{ Phys. Rev. D} \textbf{ 83} (2011) 115022,
  \href{http://dx.doi.org/10.1103/PhysRevD.83.115022}{\doi{10.1103/PhysRevD.83.115022}},
\href{http://www.arXiv.org/abs/1103.4138}{\texttt{arXiv:1103.4138}}.

\bibitem{Kang:2015nga}
\hrefCMSnoop {}{Z.~Kang, P.~Ko, and J.~Li, ``{New Physics Opportunities in the
  Boosted Di-Higgs-Boson Plus Missing Transverse Energy Signature}'',} \textit{
  Phys. Rev. Lett.} \textbf{ 116} (2016), no.~13, 131801,
  \href{http://dx.doi.org/10.1103/PhysRevLett.116.131801}{\doi{10.1103/PhysRevLett.116.131801}},
\href{http://www.arXiv.org/abs/1504.04128}{\texttt{arXiv:1504.04128}}.

\bibitem{BTV-16-002}
\hrefCMSnoop {}{{CMS Collaboration}, ``{Identification of heavy-flavour jets
  with the CMS detector in pp collisions at 13 TeV}'',} \textit{ JINST}
  \textbf{ 13} (2018), no.~05, P05011,
  \href{http://dx.doi.org/10.1088/1748-0221/13/05/P05011}{\doi{10.1088/1748-0221/13/05/P05011}},
\href{http://www.arXiv.org/abs/1712.07158}{\texttt{arXiv:1712.07158}}.

\bibitem{Chatrchyan:2013mya}
\hrefCMSnoop {}{{CMS Collaboration}, ``{Search for top squark and higgsino
  production using diphoton Higgs boson decays}'',} \textit{ Phys. Rev. Lett.}
  \textbf{ 112} (2014) 161802,
  \href{http://dx.doi.org/10.1103/PhysRevLett.112.161802}{\doi{10.1103/PhysRevLett.112.161802}},
\href{http://www.arXiv.org/abs/1312.3310}{\texttt{arXiv:1312.3310}}.

\bibitem{Khachatryan:2014mma}
\hrefCMSnoop {}{{CMS Collaboration}, ``{Searches for electroweak neutralino and
  chargino production in channels with Higgs, Z, and W bosons in pp collisions
  at 8 TeV}'',} \textit{ Phys. Rev. D} \textbf{ 90} (2014) 092007,
  \href{http://dx.doi.org/10.1103/PhysRevD.90.092007}{\doi{10.1103/PhysRevD.90.092007}},
\href{http://www.arXiv.org/abs/1409.3168}{\texttt{arXiv:1409.3168}}.

\bibitem{Aad:2012jva}
\hrefCMSnoop {}{{ATLAS Collaboration}, ``{Search for supersymmetry in events
  with photons, bottom quarks, and missing transverse momentum in proton-proton
  collisions at a centre-of-mass energy of 7 TeV with the ATLAS detector}'',}
  \textit{ Phys. Lett. B} \textbf{ 719} (2013) 261,
  \href{http://dx.doi.org/10.1016/j.physletb.2013.01.041}{\doi{10.1016/j.physletb.2013.01.041}},
\href{http://www.arXiv.org/abs/1211.1167}{\texttt{arXiv:1211.1167}}.

\bibitem{Ramond:1971gb}
\hrefCMSnoop {}{P.~Ramond, ``{Dual theory for free fermions}'',} \textit{ Phys.
  Rev. D} \textbf{ 3} (1971) 2415,
\href{http://dx.doi.org/10.1103/PhysRevD.3.2415}{\doi{10.1103/PhysRevD.3.2415}}.

\bibitem{Golfand:1971iw}
\href {http://www.jetpletters.ac.ru/ps/1584/article_24309.pdf}{Y.~A. Gol'fand
  and E.~P. Likhtman, ``Extension of the algebra of {P}oincar\'{e} group
  generators and violation of {P} invariance'',} \textit{ JETP Lett.} \textbf{
  13} (1971)
323.

\bibitem{Neveu:1971rx}
\hrefCMSnoop {}{A.~Neveu and J.~H. Schwarz, ``{Factorizable dual model of
  pions}'',} \textit{ Nucl. Phys. B} \textbf{ 31} (1971) 86,
\href{http://dx.doi.org/10.1016/0550-3213(71)90448-2}{\doi{10.1016/0550-3213(71)90448-2}}.

\bibitem{Volkov:1972jx}
\href {http://www.jetpletters.ac.ru/ps/1766/article_26864.pdf}{D.~V. Volkov and
  V.~P. Akulov, ``Possible universal neutrino interaction'',} \textit{ JETP
  Lett.} \textbf{ 16} (1972)
438.

\bibitem{Wess:1973kz}
\hrefCMSnoop {}{J.~Wess and B.~Zumino, ``A lagrangian model invariant under
  supergauge transformations'',} \textit{ Phys. Lett. B} \textbf{ 49} (1974)
  52,
\href{http://dx.doi.org/10.1016/0370-2693(74)90578-4}{\doi{10.1016/0370-2693(74)90578-4}}.

\bibitem{Wess:1974tw}
\hrefCMSnoop {}{J.~Wess and B.~Zumino, ``{Supergauge transformations in four
  dimensions}'',} \textit{ Nucl. Phys. B} \textbf{ 70} (1974) 39,
\href{http://dx.doi.org/10.1016/0550-3213(74)90355-1}{\doi{10.1016/0550-3213(74)90355-1}}.

\bibitem{Fayet:1974pd}
\hrefCMSnoop {}{P.~Fayet, ``{Supergauge invariant extension of the {H}iggs
  mechanism and a model for the electron and its neutrino}'',} \textit{ Nucl.
  Phys. B} \textbf{ 90} (1975) 104,
\href{http://dx.doi.org/10.1016/0550-3213(75)90636-7}{\doi{10.1016/0550-3213(75)90636-7}}.

\bibitem{Nilles:1983ge}
\hrefCMSnoop {}{H.~P. Nilles, ``{Supersymmetry, supergravity and particle
  physics}'',} \textit{ Phys. Rep.} \textbf{ 110} (1984) 1,
\href{http://dx.doi.org/10.1016/0370-1573(84)90008-5}{\doi{10.1016/0370-1573(84)90008-5}}.

\bibitem{Barbieri:1987fn}
\hrefCMSnoop {}{R.~Barbieri and G.~F. Giudice, ``Upper bounds on supersymmetric
  particle masses'',} \textit{ Nucl. Phys. B} \textbf{ 306} (1988) 63,
\href{http://dx.doi.org/10.1016/0550-3213(88)90171-X}{\doi{10.1016/0550-3213(88)90171-X}}.

\bibitem{Dimopoulos:1995mi}
\hrefCMSnoop {}{S.~Dimopoulos and G.~F. Giudice, ``{Naturalness constraints in
  supersymmetric theories with nonuniversal soft terms}'',} \textit{ Phys.
  Lett. B} \textbf{ 357} (1995) 573,
  \href{http://dx.doi.org/10.1016/0370-2693(95)00961-J}{\doi{10.1016/0370-2693(95)00961-J}},
\href{http://www.arXiv.org/abs/hep-ph/9507282}{\texttt{arXiv:hep-ph/9507282}}.

\bibitem{Barbieri:2009ev}
\hrefCMSnoop {}{R.~Barbieri and D.~Pappadopulo, ``{S-particles at their
  naturalness limits}'',} \textit{ JHEP} \textbf{ 10} (2009) 061,
  \href{http://dx.doi.org/10.1088/1126-6708/2009/10/061}{\doi{10.1088/1126-6708/2009/10/061}},
\href{http://www.arXiv.org/abs/0906.4546}{\texttt{arXiv:0906.4546}}.

\bibitem{Papucci:2011wy}
\hrefCMSnoop {}{M.~Papucci, J.~T. Ruderman, and A.~Weiler, ``{Natural {SUSY}
  endures}'',} \textit{ JHEP} \textbf{ 09} (2012) 035,
  \href{http://dx.doi.org/10.1007/JHEP09(2012)035}{\doi{10.1007/JHEP09(2012)035}},
\href{http://www.arXiv.org/abs/1110.6926}{\texttt{arXiv:1110.6926}}.

\bibitem{bib-sms-2}
\hrefCMSnoop {}{J.~Alwall, P.~C. Schuster, and N.~Toro, ``Simplified models for
  a first characterization of new physics at the {LHC}'',} \textit{ Phys. Rev.
  D} \textbf{ 79} (2009) 075020,
  \href{http://dx.doi.org/10.1103/PhysRevD.79.075020}{\doi{10.1103/PhysRevD.79.075020}},
\href{http://www.arXiv.org/abs/0810.3921}{\texttt{arXiv:0810.3921}}.

\bibitem{bib-sms-3}
\hrefCMSnoop {}{J.~Alwall, M.-P. Le, M.~Lisanti, and J.~G. Wacker,
  ``{Model-independent jets plus missing energy searches}'',} \textit{ Phys.
  Rev. D} \textbf{ 79} (2009) 015005,
  \href{http://dx.doi.org/10.1103/PhysRevD.79.015005}{\doi{10.1103/PhysRevD.79.015005}},
\href{http://www.arXiv.org/abs/0809.3264}{\texttt{arXiv:0809.3264}}.

\bibitem{bib-sms-4}
\hrefCMSnoop {}{D.~Alves {et~al.}, ``Simplified models for {LHC} new physics
  searches'',} \textit{ J. Phys. G} \textbf{ 39} (2012) 105005,
  \href{http://dx.doi.org/10.1088/0954-3899/39/10/105005}{\doi{10.1088/0954-3899/39/10/105005}},
\href{http://www.arXiv.org/abs/1105.2838}{\texttt{arXiv:1105.2838}}.

\bibitem{Chatrchyan:2008aa}
\hrefCMSnoop {}{{CMS Collaboration}, ``{The {CMS} experiment at the {CERN}
  {LHC}}'',} \textit{ JINST} \textbf{ 3} (2008) S08004,
\href{http://dx.doi.org/10.1088/1748-0221/3/08/S08004}{\doi{10.1088/1748-0221/3/08/S08004}}.

\bibitem{CMS-PRF-14-001}
\hrefCMSnoop {}{{CMS Collaboration}, ``Particle-flow reconstruction and global
  event description with the {CMS} detector'',} \textit{ JINST} \textbf{ 12}
  (2017) P10003,
  \href{http://dx.doi.org/10.1088/1748-0221/12/10/P10003}{\doi{10.1088/1748-0221/12/10/P10003}},
\href{http://www.arXiv.org/abs/1706.04965}{\texttt{arXiv:1706.04965}}.

\bibitem{Cacciari:2008gp}
\hrefCMSnoop {}{M.~Cacciari, G.~P. Salam, and G.~Soyez, ``The anti-$k_t$ jet
  clustering algorithm'',} \textit{ JHEP} \textbf{ 04} (2008) 063,
  \href{http://dx.doi.org/10.1088/1126-6708/2008/04/063}{\doi{10.1088/1126-6708/2008/04/063}},
\href{http://www.arXiv.org/abs/0802.1189}{\texttt{arXiv:0802.1189}}.

\bibitem{Cacciari:2011ma}
\hrefCMSnoop {}{M.~Cacciari, G.~P. Salam, and G.~Soyez, ``{FastJet} user
  manual'',} \textit{ Eur. Phys. J. C} \textbf{ 72} (2012) 1896,
  \href{http://dx.doi.org/10.1140/epjc/s10052-012-1896-2}{\doi{10.1140/epjc/s10052-012-1896-2}},
\href{http://www.arXiv.org/abs/1111.6097}{\texttt{arXiv:1111.6097}}.

\bibitem{Khachatryan:2016kdb}
\hrefCMSnoop {}{{CMS Collaboration}, ``{Jet energy scale and resolution in the
  CMS experiment in pp collisions at 8 TeV}'',} \textit{ JINST} \textbf{ 12}
  (2017) P02014,
  \href{http://dx.doi.org/10.1088/1748-0221/12/02/P02014}{\doi{10.1088/1748-0221/12/02/P02014}},
\href{http://www.arXiv.org/abs/1607.03663}{\texttt{arXiv:1607.03663}}.

\bibitem{Cacciari:2007fd}
\hrefCMSnoop {}{M.~Cacciari and G.~P. Salam, ``{Pileup subtraction using jet
  areas}'',} \textit{ Phys. Lett. B} \textbf{ 659} (2008) 119,
  \href{http://dx.doi.org/10.1016/j.physletb.2007.09.077}{\doi{10.1016/j.physletb.2007.09.077}},
\href{http://www.arXiv.org/abs/0707.1378}{\texttt{arXiv:0707.1378}}.

\bibitem{JetSub:Prune}
\hrefCMSnoop {}{S.~D. Ellis, C.~K. Vermilion, and J.~R. Walsh, ``Recombination
  algorithms and jet substructure: Pruning as a tool for heavy particle
  searches'',} \textit{ Phys. Rev. D} \textbf{ 81} (2010) 094023,
  \href{http://dx.doi.org/10.1103/PhysRevD.81.094023}{\doi{10.1103/PhysRevD.81.094023}},
  \href{http://www.arXiv.org/abs/0912.0033}{\texttt{arXiv:0912.0033}}.

\bibitem{Khachatryan:2016bia}
\hrefCMSnoop {}{{CMS Collaboration}, ``{The CMS trigger system}'',} \textit{
  JINST} \textbf{ 12} (2017) P01020,
  \href{http://dx.doi.org/10.1088/1748-0221/12/01/P01020}{\doi{10.1088/1748-0221/12/01/P01020}},
\href{http://www.arXiv.org/abs/1609.02366}{\texttt{arXiv:1609.02366}}.

\bibitem{Arnison:1983rp}
\hrefCMSnoop {}{{UA1} Collaboration, ``Experimental observation of isolated
  large transverse energy electrons with associated missing energy at
  {$\sqrt{s}= 540$~GeV}'',} \textit{ Phys. Lett. B} \textbf{ 122} (1983) 103,
\href{http://dx.doi.org/10.1016/0370-2693(83)91177-2}{\doi{10.1016/0370-2693(83)91177-2}}.

\bibitem{Alwall:2014hca}
J.~Alwall\hrefCMSnoop {}{ {et~al.}, ``{The automated computation of tree-level
  and next-to-leading order differential cross sections, and their matching to
  parton shower simulations}'',} \textit{ JHEP} \textbf{ 07} (2014) 079,
  \href{http://dx.doi.org/10.1007/JHEP07(2014)079}{\doi{10.1007/JHEP07(2014)079}},
\href{http://www.arXiv.org/abs/1405.0301}{\texttt{arXiv:1405.0301}}.

\bibitem{Ball:2014uwa}
\hrefCMSnoop {}{{NNPDF} Collaboration, ``{Parton distributions for the LHC Run
  II}'',} \textit{ JHEP} \textbf{ 04} (2015) 040,
  \href{http://dx.doi.org/10.1007/JHEP04(2015)040}{\doi{10.1007/JHEP04(2015)040}},
\href{http://www.arXiv.org/abs/1410.8849}{\texttt{arXiv:1410.8849}}.

\bibitem{Sirunyan:2017cwe}
\hrefCMSnoop {}{{CMS Collaboration}, ``{Search for supersymmetry in multijet
  events with missing transverse momentum in proton-proton collisions at 13
  TeV}'',} \textit{ Phys. Rev. D} \textbf{ 96} (2017) 032003,
  \href{http://dx.doi.org/10.1103/PhysRevD.96.032003}{\doi{10.1103/PhysRevD.96.032003}},
\href{http://www.arXiv.org/abs/1704.07781}{\texttt{arXiv:1704.07781}}.

\bibitem{Agostinelli:2002hh}
\hrefCMSnoop {}{{GEANT4} Collaboration, ``{GEANT4}---a simulation toolkit'',}
  \textit{ Nucl. Instrum. Meth. A} \textbf{ 506} (2003) 250,
\href{http://dx.doi.org/10.1016/S0168-9002(03)01368-8}{\doi{10.1016/S0168-9002(03)01368-8}}.

\bibitem{Borschensky:2014cia}
C.~Borschensky\hrefCMSnoop {}{ {et~al.}, ``{Squark and gluino production cross
  sections in pp collisions at $\sqrt{s}$ = 13, 14, 33 and 100 TeV}'',}
  \textit{ Eur. Phys. J. C} \textbf{ 74} (2014) 3174,
  \href{http://dx.doi.org/10.1140/epjc/s10052-014-3174-y}{\doi{10.1140/epjc/s10052-014-3174-y}},
\href{http://www.arXiv.org/abs/1407.5066}{\texttt{arXiv:1407.5066}}.

\bibitem{Catani:2003zt}
\hrefCMSnoop {}{S.~Catani, D.~de~Florian, M.~Grazzini, and P.~Nason, ``{Soft
  gluon resummation for Higgs boson production at hadron colliders}'',}
  \textit{ JHEP} \textbf{ 07} (2003) 028,
  \href{http://dx.doi.org/10.1088/1126-6708/2003/07/028}{\doi{10.1088/1126-6708/2003/07/028}},
\href{http://www.arXiv.org/abs/hep-ph/0306211}{\texttt{arXiv:hep-ph/0306211}}.

\bibitem{Cacciari:2003fi}
M.~Cacciari\hrefCMSnoop {}{ {et~al.}, ``The $t\bar{t}$ cross-section at {1.8
  TeV and 1.96 TeV}: a study of the systematics due to parton densities and
  scale dependence'',} \textit{ JHEP} \textbf{ 04} (2004) 068,
  \href{http://dx.doi.org/10.1088/1126-6708/2004/04/068}{\doi{10.1088/1126-6708/2004/04/068}},
\href{http://www.arXiv.org/abs/hep-ph/0303085}{\texttt{arXiv:hep-ph/0303085}}.

\bibitem{Cowan:2010js}
\hrefCMSnoop {}{G.~Cowan, K.~Cranmer, E.~Gross, and O.~Vitells, ``Asymptotic
  formulae for likelihood-based tests of new physics'',} \textit{ Eur. Phys. J.
  C} \textbf{ 71} (2011) 1554,
  \href{http://dx.doi.org/10.1140/epjc/s10052-011-1554-0}{\doi{10.1140/epjc/s10052-011-1554-0}},
  \href{http://www.arXiv.org/abs/1007.1727}{\texttt{arXiv:1007.1727}}.
[Erratum: \DOI{10.1140/epjc/s10052-013-2501-z}].

\bibitem{bib-cls}
\hrefCMSnoop {}{A.~L. Read, ``Presentation of search results: the {$CL_s$}
  technique'',} in \textit{ Durham IPPP Workshop: Advanced Statistical
  Techniques in Particle Physics}, p.~2693.
\newblock Durham, UK, March, 2002.
\newblock [J. Phys. G 28 (2002) 2693].
  \href{http://dx.doi.org/10.1088/0954-3899/28/10/313}{\doi{10.1088/0954-3899/28/10/313}}.

\bibitem{Junk1999}
\hrefCMSnoop {}{T.~Junk, ``{Confidence level computation for combining searches
  with small statistics}'',} \textit{ Nucl. Instrum. Meth. A} \textbf{ 434}
  (1999) 435,
  \href{http://dx.doi.org/10.1016/S0168-9002(99)00498-2}{\doi{10.1016/S0168-9002(99)00498-2}},
\href{http://www.arXiv.org/abs/hep-ex/9902006}{\texttt{arXiv:hep-ex/9902006}}.

\bibitem{cms-note-2011-005}
\href {https://cds.cern.ch/record/1379837}{{ATLAS and CMS Collaborations, LHC
  Higgs Combination Group}, ``{Procedure for the LHC Higgs boson search
  combination in Summer 2011}'',} Technical Report CMS-NOTE-2011-005,
  ATL-PHYS-PUB-2011-11, CERN, 2011.

\end{thebibliography}\endgroup

\cleardoublepage \appendix\section{The CMS Collaboration \label{app:collab}}\begin{sloppypar}\hyphenpenalty=5000\widowpenalty=500\clubpenalty=5000\vskip\cmsinstskip
\textbf{Yerevan Physics Institute,  Yerevan,  Armenia}\\*[0pt]
A.M.~Sirunyan,  A.~Tumasyan
\vskip\cmsinstskip
\textbf{Institut f\"{u}r Hochenergiephysik,  Wien,  Austria}\\*[0pt]
W.~Adam,  F.~Ambrogi,  E.~Asilar,  T.~Bergauer,  J.~Brandstetter,  E.~Brondolin,  M.~Dragicevic,  J.~Er\"{o},  A.~Escalante Del Valle,  M.~Flechl,  M.~Friedl,  R.~Fr\"{u}hwirth\cmsAuthorMark{1},  V.M.~Ghete,  J.~Grossmann,  J.~Hrubec,  M.~Jeitler\cmsAuthorMark{1},  A.~K\"{o}nig,  N.~Krammer,  I.~Kr\"{a}tschmer,  D.~Liko,  T.~Madlener,  I.~Mikulec,  E.~Pree,  N.~Rad,  H.~Rohringer,  J.~Schieck\cmsAuthorMark{1},  R.~Sch\"{o}fbeck,  M.~Spanring,  D.~Spitzbart,  A.~Taurok,  W.~Waltenberger,  J.~Wittmann,  C.-E.~Wulz\cmsAuthorMark{1},  M.~Zarucki
% ...
\vskip\cmsinstskip
% ...
1:~Also at Vienna University of Technology,  Vienna,  Austria\\
% ...
\end{sloppypar}
\end{document}